\documentstyle[12pt]{article}
\title{Stochastic stability in spatial three-player games} 
\author{Jacek Mi\c{e}kisz \\ Institute of Applied Mathematics \\
and Mechanics \\ Warsaw University  \\ ul. Banacha 2  \\ 02-097
Warsaw, Poland 
\\ e-mail: miekisz@mimuw.edu.pl} 
\begin{document} 
\baselineskip=20pt
\maketitle 

\newtheorem{theo}{Theorem}
\newtheorem{defi}{Definition}
\newtheorem{prop}{Proposition}

\noindent {\bf Abstract}: We discuss long-run behavior of stochastic dynamics
of many interacting agents. In particular, three-player spatial games are studied. 
The effect of the number of players and the noise level
on the stochastic stability of Nash equilibria is investigated.
\vspace{3mm}

\vspace{3mm}


\eject

\section{Introduction}
Socio-economic systems can be viewed as systems of many interacting 
agents or players (see for example Santa Fe collection of papers on economic 
complex systems \cite{santa} and econophysics papers on Minority Game \cite{ekono}). 
We may then try to derive their global behavior from individual interactions 
between their basic entities. Such approach is fundamental in statistical physics 
which deals with systems of many interacting particles. We will explore similarities 
and differences between systems of many interacting players maximizing their individual 
payoffs and particles minimizing their interaction energy. 

We will consider here game-theoretic models of many interacting agents \cite{wei,hof}. 
In such models, agents have at their disposal certain strategies and their payoffs 
in a game depend on strategies chosen both by them and by their opponents.
In spatial games, agents are located on vertices of certain graphs and they interact
only with their neighbors \cite{blume1,ellis1,ellis2,nowak1,linnor,doebeli,
sabo,hauert}. The central concept in game theory is that of a Nash equilibrium. 
A configuration of strategies (an assignment of strategies to agents) 
is a Nash equilibrium, if no agent, for fixed strategies of his opponents,  
can increase his payoff by deviating from his current strategy. 
In spatial models, a Nash equilibrium is called a Nash configuration.
We see that the notion of a Nash configuration is similar to that 
of a ground-state configuration in systems of interacting particles.

In most models with many players, their strategic interaction 
is decomposed into a sum of two-player games. Only recently 
there have appeared some systematic studies of truly multi-player games 
\cite{broom,kim,bukmie}. Here we consider spatial games with players located 
on vertices of the triangular lattice. Each agent plays six three-player games 
with his neighbors on the same triangle.

One of the fundamental problems in game theory is that of the equilibrium selection
in games with multiple Nash equilibria. We will discuss here the dynamic approach 
to that problem. It may happen that only some equilibria are asymptotically stable
in some specific dynamics. We will be concerned here with a particular stochastic dynamics. 
Namely, at discrete moments of time, a randomly chosen player may change his strategy. 
He adopts with a high probability a strategy which is the best response
to strategies of his neighbors, that is a strategy that maximizes 
the sum of the payoffs of individual games, and with a small probability,
representing the noise of the system, he makes a ``mistake''. 
Such process is repeated infinitely many times. To describe the long-run behavior
of stochastic dynamics, Foster and Young \cite{foya} introduced a concept 
of stochastic stability. A configuration of strategies is {\bf stochastically stable} 
if it has a positive probability in the stationary state of the above dynamics 
in the zero-noise limit, that is the zero probability of mistakes. 
It means that in the long run we observe it with a positive frequency.
However, for any arbitrarily low but fixed noise, if the number 
of players is big enough, the probability of any individual configuration 
is practically zero. It means that for a large number of
players, to observe a stochastically stable configuration we must assume that players 
make mistakes with extremely small probabilities. However, as indicated 
by van Damme and Weibull \cite{damwei}, a small probability of mistakes 
should involve some some cost of learning strategies played by neighbors. 
To avoid paying these prohibitively big costs, players settle for regimes 
with low but not extremely low noise. On the other hand, it may happen that 
in the long run, for a low but fixed noise and sufficiently big number 
of players, the stationary state is highly concentrated on an ensemble
consisting of one Nash configuration and its small perturbations, 
i.e. configurations, where most players play the same strategy.
We will call such configurations {\bf low-noise ensemble stable.}

We will investigate here the effect of the noise level and the number of players
on their long-run behavior. In the first part of our paper we will consider 
the so-called potential games \cite{mon}. In such games, if any single player 
changes his strategy, then the payoff differences are the same for all players. 
This is in absolute analogy to systems of interacting particles, 
where instead of maximizing payoffs, particles minimize their interaction energy. 
We will exploit this analogy to describe long-run behavior of potential three-player games 
with two Nash configurations. We will show that a configuration  
can be stochastically stable but nevertheless may appear in the long run 
with an arbitrarily small frequency if the number of players is large enough -  
it is not low-noise ensemble stable. In the second part of our paper 
we will present an example of a simple nonpotential three-player spatial game, 
where stochastic stability depends on the number of players.  

In Section 2, we introduce spatial three-player games. In Section 3,
we compare stochastic and ensemble stability in potential games.
In Section 4, we discuss nonpotential games. Discussion follows in Section 5.
 
\section{Spatial three-player games}

\noindent Let $\Lambda$ be a finite subset of the triangular lattice. 
Every site of $\Lambda$ is occupied by one player who
has at his disposal one of two different strategies. 
Let $S=\{A,B\}$ be the set of strategies, then $\Omega_{\Lambda}=S^{\Lambda}$ is the space
of all possible configurations of players. For every $i \in \Lambda$, 
$X_{i}$ is the strategy of the $i-$th player in the configuration 
$X \in \Omega_{\Lambda}$ and $X_{-i}$
denotes strategies of all remaining players; $X$ therefore can be represented 
as the pair $(X_{i},X_{-i})$. $U: S \times S \times S \rightarrow R$ 
is a payoff function of our game. Without loss of generality (see a discussion below)
it can be represented by two matrices: 

\begin{equation}
\begin{array}{c}
U=\left( \left( 
\begin{array}{cc}
a & 0 \\ 
0 & b
\end{array}
\right) ,\left( 
\begin{array}{cc}
0 & 0 \\ 
b & c
\end{array}
\right) \right) 
\end{array},
\end{equation}

\noindent where the $ij$ entry, $i, j =A, B$, of the first matrix
is the payoff of the first (row) player when he plays strategy $i$, 
the second (column) player plays the strategy $j$ and the third 
(matrix) player plays the strategy A; the second matrix represents payoffs
of the first player when the third player plays the strategy B.
We assume that all players are the same and hence payoffs 
of a column and a matrix player can be easily deduced from the above matrices;
such games are called symmetric. 

Every player interacts only with his nearest neighbors and his payoff 
is the sum of the payoffs resulting from individual games 
(six games on the triangular lattice).
We assume that he has to use the same strategy for all neighbors. 
For $X \in \Omega_{\Lambda}$ we denote by $\nu_{i}(X)$ the payoff 
of the $i-$th player in the configuration $X$:
\begin{equation}
\nu_{i}(X)=\sum_{(j, k)}U(X_{i}, X_{j}, X_{k}),
\end{equation}
where the summation is with respect to six elementary triangles containing $i$.
\begin{defi}
$X \in \Omega_{\Lambda}$ is a {\bf Nash configuration} if for every $i \in \Lambda$
and $Y_{i} \in S$, $\nu_{i}(X_{i},X_{-i}) \geq \nu_{i}(Y_{i},X_{-i})$.
\end{defi}
Let us note that the notion of a Nash configuration involves not only  payoff
functions but also the spatial structure of players. It is similar to the notion 
of a ground-state configuration in classical lattice-gas models of interacting particles.
However, there are differences. One cannot decrease the energy of a ground 
state-configuration by any local change of particles.  From this follows
the existence of a ground-state configuration for any model with finite-range interactions.
In the definition of a Nash configuration we are allowed to make only one-site changes.
As a consequence of this restriction, a Nash configuration may not exist.
We will be not concerned here with such situations.

Let us notice that if $a>0$ and $c>0$, then there are two homogeneous Nash configurations:
$X^{A}$ and $X^{B}$, where all players play the same strategy, $A$ or $B$ respectively.
If  $a>0$, $c<0$ and $b>0$, then we have a homogeneous Nash configuration $X^{A}$ 
and three configurations, related by translations, where on every elementary triangle 
there are two $B$ players and one $A$ player. We donote by $X^{ABB}$ one of these 
configurations. We see that for above payoff parameters, there are multiple Nash 
configurations. We are therefore faced with a standard game-theoretic 
problem of equilibrium selection. In the following, we will discuss one of the dynamics 
used in evolutionary game theory.  

We start with the deterministic dynamics of the {\bf best-response rule}. 
Namely, at each discrete moment of time $t=1,2,...$, a randomly chosen player may update 
his strategy. He simply adopts the strategy, $X_{i}^{t}$, which gives him 
the maximal total payoff $\nu_{i}(X_{i}^{t}, X^{t-1}_{-i})$ 
for given $X^{t-1}_{-i}$, a configuration of strategies 
of remaining players at time $t-1$. 

Now we allow players to make mistakes with a small probability, 
that is to say they may not choose the best response. 
It is reasonably to expect that the probability of making an error
should increase if payoffs from alternate strategies approach 
the payoff of the best-response strategy. We will consider here 
a well-known in statistical mechanics exponential 
rule which is used in game-theoretic and economic literature 
under the name of the log-linear rule \cite{blume1,young2}. 
We assume that the probability of chosing by the $i-$th player 
the strategy $X_{i}^{t}$ at time $t$ is given by the following 
conditional probability:

\begin{equation}
p_{i}^{\beta}(X_{i}^{t}|X_{-i}^{t-1})=
\frac{e^{\beta\nu_{i}( X_{i}^{t},X_{-i}^{t-1})}}{\sum_{Y_{i} \in S}
e^{\beta\nu_{i}(Y_{i},X_{-i}^{t-1})}},
\end{equation}
where $1/\beta >0$ measures the noise level.

Let us observe that if $\beta \rightarrow \infty$, 
$p_{i}^{\beta}$ converges pointwise to the best-response rule.
Such stochastic dynamics is an example of an irreducible 
Markov chain with $|S^{\Lambda}|$ states (there is a nonzero probability 
of transition from any state to any other state in finite number of steps). 
Therefore, it has the unique stationary probability distribution 
(also called a stationary state) denoted by 
$\mu_{\Lambda}^{\beta}.$  The following definition was introduced 
by Foster and Young \cite{foya}:

\begin{defi}
$X \in \Omega_{\Lambda}$ is {\bf stochastically stable} 
if $\lim_{\beta \rightarrow \infty}\mu_{\Lambda}^{\beta}(X) >0.$
\end{defi} 
If $X$ is stochastically stable, then the frequency of visiting $X$ converges to
a positive number along any time trajectory almost surely. 

Stationary distributions of log-linear dynamics can be explicitly constructed 
for the class of the so-called potential games \cite{mon,young2}.
In such games, if any single player changes his strategy, 
then the payoff differences are the same for all players.
More precisely, a game is a {\bf potential game}
if there exists a function $\rho: S \times S \times S \rightarrow R$,
invariant under any permutation of arguments such that 
for all $x,x',y,z \in S$
\begin{equation}
U(x',y,z)-U(x,y,z)=\rho(x',y,z)-\rho(x,y,z)
\end{equation}

We call this function a potential of the game.

It is easy to see that $\rho(A,A,A)=a, \rho(A,A,B)=0, \rho(A,B,B)=b,
\rho(B,B,B)=b+c$ is a potential of an elementary three-player game defined in (1).
For players on the triangular lattice playing six elementary games, 
for any $X \in \Omega_{\Lambda}$,
\begin{equation}
\rho(X)=\sum_{(i,j,k) \in \triangle \subset
\Lambda} \rho(X_{i},X_{j},X_{k}),
\end{equation}
is a potential of the configuration $X$,
where a sum is taken with respect to all elementary triangles in $\Lambda$.
We have to stress here that even if an elementary game has a potential 
it does not necessarily mean that a resulting spatial game has a potential.
This depends upon the spatial structure of interactions as we will see in Section 4.

We will now show that the following probability distribution is
the unique stationary state of our spatial game. 
\vspace{2mm}

\noindent {\bf Proposition}

$$
\mu^{\beta}_{\Lambda}(X)=\frac{e^{\beta\sum_{(i,j,k) \in \triangle}\rho(X_{i},X_{j},X_{k})}}
{\sum_{Z \in \Omega_{\Lambda}}e^{\beta\sum_{(i,j,k) \in \triangle}\rho(Z_{i},Z_{j},Z_{k})}},
$$

is the stationary state of a three-player game on the triangular lattice.
\vspace{2mm}

\noindent {\bf Proof:}

We will show that $\mu^{\beta}_{\Lambda}$ satisfies the detailed balance condition
\begin{equation}
\mu^{\beta}_{\Lambda}(X)P(X,Y)=\mu^{\beta}_{\Lambda}(Y)P(Y,X)
\end{equation}
for all $X,Y \in \Omega_{\Lambda}$, where $P(X,Y)$ is the transition probability
from $X$ to $Y$ given in (3).

Then it follows that $\mu^{\beta}_{\Lambda}$ is a stationary distribution
because
$$ \sum_{X \in \Omega_{\Lambda}}\mu^{\beta}_{\Lambda}(X)P(X,Y)=
\sum_{X \in \Omega_{\Lambda}}\mu^{\beta}_{\Lambda}(Y)P(Y,X)=\mu^{\beta}_{\Lambda}(Y)
\sum_{X \in \Omega_{\Lambda}}P(Y,X)= \mu^{\beta}_{\Lambda}(Y).$$

Let us observe that $Y$ can be different at at most one lattice site, say $i$,
which was chosen randomly (with probability  $1/|\Lambda|$) out of $\Lambda$. Let 
$$D=\frac{1}{|\Lambda|\sum_{Z \in \Omega_{\Lambda}}e^{\beta \rho(Z)}\sum_{Z_{i}\in S}
e^{\beta \sum_{(j,k)} U(Z_{i},X_{j},X_{k})}}.$$

We have

$$\mu^{\beta}_{\Lambda}(X)P(X,Y)=De^{\beta \sum_{(i,j,k) \in \triangle} \rho(X_{i},X_{j},X_{k})}
e^{\beta \sum_{(j,k)}U(Y_{i},X_{j},X_{k})}$$
$$=De^{\beta \sum_{(i,j,k) \in \triangle} \rho(X_{i},X_{j},X_{k})}
e^{\beta \sum_{(j,k)}U(X_{i},X_{j},X_{k})-\rho(X_{i},X_{j},X_{k})+\rho(Y_{i},X_{j},X_{k})}$$
$$=De^{\beta \sum_{(i,j,k) \in \triangle} \rho(Y_{i},Y_{j},Y_{k})}
e^{\beta \sum_{(j,k)}U(X_{i},X_{j},X_{k})}=  \mu^{\beta}_{\Lambda}(Y)P(Y,X)$$

so $\mu^{\beta}_{\Lambda}$ satisfies the detailed balance condition which
completes the proof of the proposition.
\vspace{2mm}

$\mu^{\beta}_{\Lambda}$ is a so-called finite-volume Gibbs state -
a probability distribution describing the equilibrium behavior of systems 
of many interacting particles. In the following section, we will investigate 
the stochastic stability of Nash configurations for different payoff parameters 
of three-player games.

\section{Stochastic and ensemble stability}
Different Nash configurations of a given game usually have different 
values of a potential. It follows from the explicit form of the stationary state 
in the Proposition that Nash configurations with the maximal potential 
are stochastically stable. We obtain immediately the following theorems.
\begin{theo} 
Let $a,c>0.$ If $a>b+c$, then $X^{A}$ is stochastically stable; if 
$a<b+c$, then $X^{B}$ is stochastically stable.  
\end{theo}

Let us notice that in our case, stochastically stable configurations appear 
in the long run with the probability $1$ in the zero-noise limit.

If $a=b+c$, then both $X^{A}$ and $X^{B}$ are stochastically stable
and in the limit of zero noise they occur with the probability $1/2$.

\begin{theo} 
Let $a>0$, $c<0$, and $b>0$. If $a>b$, then $X^{A}$ is stochastically stable; if  
$a<b$, then $X^{ABB}$ and its two translates are stochastically stable.
\end{theo}

If $a=b$, then all four Nash configurations are stochastically stable and
they occur with the probability $1/4$.

We see that if there are two or more Nash configurations
with the maximal potential, then the problem of equilibrium 
selection is still not resolved.

Let us notice that 
$\lim_{\Lambda \rightarrow {\bf Z}^{2}}\mu_{\Lambda}^{\beta}(X)=0$ 
for every $X \in S^{L}$, where $L$ is the infinite triangular lattice.
Hence for large $\Lambda$ and any nonzero nisewe may only observe,
with reasonable positive frequencies, ensembles of configurations 
and not particular configurations. It may happen that the stationary state 
is highly concentrated on an ensemble consisting of one Nash configuration 
and its small perturbations, i.e. configurations, where most players play 
the same strategy. We will call such configurations low-noise ensemble stable.

\begin{defi}
{\em $X \in \Omega_{\Lambda}$ is} {\bf low-noise ensemble stable}
{\em if for every $\epsilon> 0$, there exists $\beta(\epsilon)$ such that
for every $\beta > \beta(\epsilon)$ there exists $\Lambda(\beta)$
such that $\mu_{\Lambda}^{\beta}(Y \in \Omega_{\Lambda};Y_{i} \neq X_{i}) < \epsilon$
for any $i \in \Lambda$ if $\Lambda(\beta) \subset \Lambda$.}
\end{defi}

If $X$ is low-noise ensemble stable, then the ensemble consisting of $X$
and configurations which are different from $X$ at few sites has 
probability close to one in the stationary distribution. It may happen 
that only one of many stochastically stable Nash configurations is 
low-noise ensemble stable. We will show that this is exactly the case
of three-player games with certain payoff parameters. 

We will first consider the case of $a,c>0$, $b<0$, $a=b+c$
and therefore $a<c$.
 
We perform first the limit $\Lambda \rightarrow {\bf Z}^{2}$
and obtain a so-called infinite-volume Gibbs state

\begin{equation}
\mu^{\beta} = \lim_{\Lambda \rightarrow {\bf Z}^{2}}\mu_{\Lambda}^{\beta}
\end{equation}

We may then apply a technique developed by Bricmont and Slawny \cite{brsl1,brsl2}.
They studied low-temperature stability of the so-called dominant 
ground-state configurations. It follows directly from Theorem A in \cite{brsl2} that

\begin{equation}
\mu^{\beta}(X_{i}=A)>1-\epsilon(\beta)  
\end{equation}
for any site $i$ of the lattice and $\epsilon(\beta) \rightarrow 0$ 
as $\beta \rightarrow \infty$. 
For $b>0$ so $a>c$ we have the analogous inequalityfor the strategy $B$.
The following theorem is a simple consequence of above inequalities.
\begin{theo}
Let $a=b+c.$ If $b<0$, then $X^{A}$ is low-noise ensemble stable
and if $b>0$, then $X^{B}$ is low-noise ensemble stable. 
\end{theo} 

Theorems 1 and 3 say that for any low but fixed level of noise and $b<0$, 
if the number of players is big enough, then in the long run, almost all players 
use $A$ strategy. On the other hand, if for any fixed number of players, 
the noise level is lowered substantially, then both strategies appear with frequencies 
close to $1/2$.

Let us sketch briefly the reason of such a behavior. We assume that $a<c.$
While it is true that both Nash configurations have the same potential 
which is one-third of the payoff of the whole system (it plays the role 
of the total energy of a system of interacting particles), 
the $X^{A}$ Nash configuration has more lowest-cost excitations. 
Namely, if one player changes his strategy to $B$, then 
the potential of the configuration decreases by $6a$. 
If one player in the $X^{B}$ Nash configuration changes his strategy to $A$, 
the potential of the configuration decreases by $6c>6a$. 
Now, the probability of the occurrence of any configuration in the Gibbs state 
(which is the stationary distribution of our stochastic dynamics) 
depends on the total payoff in an exponential way. 
One then proves that the probability of an ensemble consisting of the $X^{A}$ Nash 
configuration and configurations which are different from it 
at few sites only is much bigger than the probability of the analogous
$X^{B}$-ensemble. On the other hand, configurations which are outside 
$X^{A}$ and $X^{B}$-ensembles appear with exponentially small probabilities. 
It means that for large enough systems 
(and small but not extremely small noise level) we observe in the stationary distribution 
the $X^{A}$ Nash configuration with perhaps few different strategies. 
The above argument was made into a rigorous proof for infinite systems 
of corresponding lattice-gas models of interacting particles 
by Bricmont and Slawny in \cite{brsl1,brsl2}. They would call $X^{A}$  
a dominant ground-state configuration. 

We have an analogous theorem for the other class of three-player games.

\begin{theo}
For $a>0$, $c<0$, and $a=b$, if $a<|c|$, then $X^{A}$ is low-noise ensemble stable; 
if $a>|c|$, then $X^{ABB}$ and its translates are low-noise ensemble stable.
\end{theo}   

Here the lowest-cost excitation from $X^{A}$ is still $6a$.
Let us describe the lowest-cost excitations from $X^{ABB}$.
When $B$ changes to $A$, then the payoff of the configuration 
decreases by $6b=6a$. However, if $A$ changes to $B$, the payoff
decreases by $|c|$. Therefore, if $a>|c|$, then $X^{ABB}$
has more lowest-cost excitations and hence is low-noise ensemble stable.

\section{Nonpotential three-player games}

Now we will consider an example of a three-player spatial game without
a potential. Players are now placed on a finite subset of the one-dimensional 
regular lattice ${\bf Z}$ (for simplicity we will assume periodic boundary conditions 
and therefore agents will reside on a circle). Every agent can play only one three-player game
with his right and left nearest neighbor. Although any single
game with a payoff matrix given in (1) has a potential as before but a sum 
of three-player interactions is not a potential of the spatial game. 
The reason for this is that if any agent chooses a best-response strategy,
he does not take into account a game with two left or two right neighbors.
However, his action may change their payoffs as a result of two additional 
three-player games. Hence, $\mu^{\beta}_{\Lambda}$ given in the Proposition 
is no longer a stationary state of our stochastic dynamics. 
To find stochastically stable configurations, we must resort to different methods. 
We will use the following tree representation of stationary states 
of irreducible Markov chains \cite{freiwen}. Let $(\Omega,P)$ be an irreducible Markov chain 
with a finite state space $\Omega$ and the transition probabilities given by 
$P: \Omega \times \Omega \rightarrow [0,1]$. 
Let us denote by $\mu$ its unique stationary distribution. 
For $X \in \Omega$, let an X-tree be a directed graph 
on $\Omega$ such that from every $Y \neq X$ there is a unique path to $X$
and there are no outcoming edges out of $X$. Denote by $T(X)$ the set of all X-trees
and let 
\begin{equation}
q(X)=\sum_{d \in T(X)} \prod_{(Y,Z) \in d}P(Y,Z),
\end{equation}
where the product is with respect to all edges of $d$.
The following representation of a stationary distribution $\mu$ 
was provided by Freidlin and Wentzell \cite{freiwen}:
\begin{equation}
\mu(X)=\frac{q(X)}{\sum_{Y \in \Omega}q(Y)}
\end{equation}
for all $X \in \Omega.$

We will now use the above characterisation of a stationary distribution
to find stochastically stable states in our nonpotential game for the case
of $a,c>0$.

Let us note that $X^{A}$ and $X^{B}$ are the only absorbing states 
of the noise-free dynamics. When we start with any state different 
from $X^{A}$ and $X^{B}$, then after a finite number of steps 
of the best-response dynamics we arrive at either 
$X^{A}$ or $X^{B}$ and then stay there forever. 
It follows from the above tree representation of the stationary distribution that 
any state different from $X^{A}$ and $X^{B}$ has zero probability in the zero-noise limit. 
Moreover, in order to study the zero-noise limit of the stationary distribution, 
it is enough to consider paths between absorbing states. More precisely, 
we construct X-trees with absorbing states as vertices. 
The family of such trees is denoted by $\tilde{T}(X)$  Let  
\begin{equation}
q_{m}(X)=max_{d \in \tilde{T}(X)} \prod_{(Y,Z) \in d}\tilde{P}(Y,Z),
\end{equation}
where $\tilde{P}(Y,Z)= max \prod_{(W,W')}P(W,W')$,
and the last product is taken along any path joining $Y$ with $Z$ on the full graph 
and the maximum is taken with respect to all such paths. 

Now we may observe that in our three-player game, 
if $lim_{\beta \rightarrow \infty} q_{m}(X^{B})/q_{m}(X^{A})=0$, 
then $X^{A}$ is stochastically stable. Therefore we have to compare 
trees with biggest $q_{m}$ in (11); such trees we call maximal.

Now we will use the above tree representation of a stationary state 
in two different noise models. We begin with a stochastic dynamics 
with a state-independent noise. Namely, at each discrete moment of time, 
a randomly chosen agent plays the best response with the probability $1-\epsilon$ 
and with the probability $\epsilon$ he makes a mistake. Below we assume that $a,c>0$
so there are two Nash configurations, $X^{A}$ and $X^{B}.$ 

\begin{theo}
For the state-independent noise, if $b<0$, then $X^{A}$ is stochastically stable, 
if $b>0$, then  $X^{B}$ is stochastically stable.
\end{theo}

\noindent {\bf Proof}: The theorem follows from the observation that 
if $b<0$, then $q_{m}(X^{A})$ is of order $\epsilon$ and $q_{m}(X^{B})$ 
is of order $\epsilon^{|\Lambda|/2}$, and if $b>0$, then it is the other way around. 
\vspace{3mm}

Now we come back to the state-dependent log-linear noise.
\begin{theo}
For the log-linear noise, if $a<c$, then for every small $b<0$, there is $K(b)$ such that
$X^{A}$ is stochastically stable if $|\Lambda| >K(b)$ and $X^{B}$ 
is stochastically stable if $|\Lambda| < K(b)$. 
\end{theo}
{\bf Proof}: If $|b|<a$, then we get
\begin{equation}
q_{m}(X^{A})=\frac{1}{(1+e^{\beta c})(1+e^{\beta b})^{|\Lambda|-2}(1+e^{-\beta a})},
\end{equation}
\begin{equation}
q_{m}(X^{B})=\frac{1}{(1+e^{\beta a})(1+e^{-\beta b})^{|\Lambda|-2}(1+e^{-\beta c})}.
\end{equation}  
We also have that $\lim_{b \rightarrow 0}K(b) = \infty$.

For $a>c$ and $b<0$, it follows from the above expressions of $q_{m}(X^{A})$ 
and $q_{m}(X^{A})$ that $X^{A}$ is stochastically stable for any number of players. 
We see that in nonpotential games stochastic stability may depend upon 
the number of players. Let us notice that for any arbitrarily large $c$ and $b<0$, 
if the number of players is sufficiently big, then in the zero-noise limit, 
all of them play the inefficient strategy $A$ which gives them the lower payoff 
than the strategy $B$ in the configuration $X^{B}$.  

\section{Summary}

To address the problem of equilibrium selection 
in spatial games with many players, we introduced 
the concept of low-noise ensemble stability.
We showed that in certain symmetric three-player games with two strategies,
there exist Nash configurations that are stochastically stable
but not low-noise ensemble stable. It means that for any
arbitrarily low but fixed noise, if the number of players is large enough,
then some stochastically stable strategies are played with arbitrarily small
frequencies. We also showed that for nonpotential three-player games, 
stochastic stability may depend upon the number of players.
\vspace{3mm}

\noindent {\bf Acknowledgments} I would like to thank the Polish Committee for
Scientific Research, for a financial support under the grant KBN 5 P03A 025 20.


\vspace{3mm}

\begin{thebibliography}{99}

\bibitem{santa}{The Economy as an Evolving Complex System II, 
W. B. Arthur, S. N. Durlauf, and D. A. Lane, eds. (Addison-Wesley, Reading, MA, 1997).}
\bibitem{ekono}{Econophysics bulletin on www.unifr.ch/econophysics}

\bibitem{wei}{J. Weibull, Evolutionary Game Theory 
(MIT Press, Cambridge MA, 1995).}
\bibitem{hof}{J. Hofbauer and K. Sigmund, Evolutionary Games 
and Population Dynamics (Cambridge University
Press, Cambridge, 1998).}

\bibitem{blume1}{L. E. Blume, Games Econ. Behav. 5 (1993) 387.} 
\bibitem{ellis1}{G. Ellison, Econometrica 61 (1993) 1047.} 
\bibitem{ellis2}{G. Ellison, Review of Economic Studies 67 (2000) 17.}

\bibitem{nowak1}{M. A. Nowak and R. M. May, International Journal 
of Bifurcation and Chaos 3 (1993) 35.}

\bibitem{linnor}{K. Lindgren and M. G. Nordahl, Physica D 75 (1994) 292.} 
\bibitem{doebeli}{K. Brauchli, T. Killingback, and M. Doebeli, 
Journal of Theoretical Biology 200 (1999) 405.}
\bibitem{sabo}{G. Szab\'{o}, T. Antal, P. Szab\'{o}, and M. Droz, 
Phys. Rev. E 62 (2000) 1095.}
\bibitem{hauert}{Ch. Hauert, International Journal 
of Bifurcation and Chaos 12 (2002) 1531.} 

\bibitem{broom}{M. Broom, C. Cannings and G. T. Vickers, Bull. Math. Biology 59 (1997) 931.}
\bibitem{kim}{Y. Kim, Games Econ. Behav. 15 (1996) 203.}
\bibitem{bukmie}{M. Bukowski and J. Mi\c{e}kisz, to appear in Int. J. Game Theory (2004), 
www.mimuw.edu.pl/$\sim$miekisz/multi.ps}

\bibitem{foya}{D. Foster and P. H. Young, Theoretical Population Biology 38 (1990) 219.} 

\bibitem{damwei}{E. van Damme and J. Weibull, J. Econ. Theory 106 (2002) 296.}

\bibitem{mon}{D Monderer and L. S. Shapley, Games Econ. Behav. 14 (1996) 124.}

\bibitem{young2}{H. P. Young, Individual Strategy and Social
Structure: An Evolutionary Theory of Institutions (Princeton
University Press, Princeton, 1998).}

\bibitem{brsl1}{J. Bricmont and J. Slawny,  First order phase transitions 
and perturbation theory in Statistical Mechanics and Field Theory: Mathematical Aspects,
(Lecture Notes in Physics 257. Springer-Verlag, 1986).} 
\bibitem{brsl2}{J. Bricmont and J. Slawny, J. Stat. Phys. 54 (1989) 89.}

\bibitem{freiwen}{M. Freidlin and A. Wentzell,  
Random Perturbations of Dynamical Systems (Springer Verlag, New York 1984).}

\end{thebibliography}
\end{document}